\documentclass[twocolumn]{autart}

\usepackage{tikz}
\usepackage{latexsym, amssymb}
\usepackage{graphicx}
\usepackage{amsmath, amsbsy}
\usepackage{amsopn, amstext}
\usepackage{hyperref}
\usepackage{cancel, color}
\usepackage{epstopdf}

\DeclareMathOperator{\Col}{Col}
\DeclareMathOperator{\Row}{Row}

\DeclareMathOperator{\lcm}{lcm}

\def\cal{\mathcal}

\def\ra{\rightarrow}

\def\d{\delta}

\def\0{{\bf 0}}

\def\dsum{\mathop{\sum}\limits}
\def\thefootnote{*}

\newtheorem{dfn}[thm]{Definition}
\newtheorem{prp}[thm]{Proposition}
\newtheorem{exa}[thm]{Example}

\begin{document}

\begin{frontmatter}

\title{Analysis of Discrete-Time Switched Linear Systems under Logic Dynamic Switching\thanksref{footnoteinfo}} 

\thanks[footnoteinfo]{This work was partially supported by Shanghai Pujiang Program under  Grant 21PJ1413100, the National Natural Science Foundation of China under Grant 62103305, Shanghai Municipal Science and Technology Major Project under grant 2021SHZDZX0100, and China Postdoctoral Science Foundation 2021M703423 and 2022T150686.}

\author[1,2]{Xiao Zhang}\ead{xiaozhang@amss.ac.cn},    
\author[3,4]{Min Meng}\ead{mengmin@tongji.edu.cn}, 
\author[1,5]{Zhengping Ji}\ead{jizhengping@amss.ac.cn} 
\address[1]{Key Laboratory of Systems and Control, Academy of Mathematics and Systems Science, Chinese Academy of Sciences, Beijing 100190, P.R. China}             
\address[2]{National Center for Mathematics and Interdisciplinary Sciences, Chinese Academy of Sciences, Beijing 100190, P.R. China}
\address[3]{Department of Control Science and Engineering,
Tongji University, Shanghai 201804, P.R. China} 
\address[4]{Shanghai Research Institute of Intelligent Autonomous Systems, Tongji University, Shanghai 201210, P.R. China}
\address[5]{School of Mathematical Sciences, University of Chinese Academy of Sciences, Beijing 100049, P.R. China}


\begin{abstract}                          
The control properties of discrete-time switched linear systems (SLS) with switching signals generated by logical dynamic systems are studied using the semi-tensor product (STP) approach. With the algebraic state space representation (ASSR), the linear modes and the logical generators are aggregated as a hybrid system, leading to the criteria of reachability, controllability, observability, and reconstructibility of the SLSs. Algorithms for checking these properties are given. Then, two kinds of realization problems concerning whether the logical dynamic systems can generate the desired switching signals are investigated, and necessary and sufficient conditions for the realisability of the required switching signals are given with respect to the cases of fixed operating time switching and finite reference signal switching.
\end{abstract}

\end{frontmatter}

\section{Introduction}
A switched linear system (SLS) is a dynamical system consisting of a finite number of linear subsystems (or modes) and a logical rule that controls the activating subsystem during operation time. SLSs can be either continuous-time or discrete-time and can be discretized if the unit switching time is known. Due to the flexibility of mode selection, SLSs offer superior performance and can perform a wider range of tasks than any single subsystem, making them a popular choice for analysis and control. In the last three decades, there have been many advances in the study of SLSs, including stability and stabilization \cite{fen02,che05,gur07,lin09,sun09,kun15,guo22}, reachability and controllability \cite{ge01,sun01,sun02,xie02,ji08}, observability and reconstructibility \cite{ge01,bag07,gom10,tan13,kus18}, and more.

In most of the existing literature on SLSs, the logical rules that generate the switching signals are usually freely chosen or determined by the range of the physical state $x(t)$, i.e. the rules are given as piecewise constant maps from each switching time to the index set of the subsystems or state-feedback switching (for example, the mixed logical-dynamical systems \cite{bem99}), rather than as outputs of some logical-dynamical processes. This means that one cannot consider an SLS as a true hybrid system containing both physical and logical states, but must split it into regulatory and operational parts. Perhaps this is not surprising given the following facts:
\begin{itemize}
    \item It is well known that modeling logical dynamical systems is not an easy task.
    \item Assuming that the logical rules are somehow modeled, it is also meaningless if you cannot aggregate the obtained model with the linear subsystems. 
\end{itemize}
However, in SLSs where the switching is generated by logical dynamic systems, we can use the semi-tensor product (STP) of matrices to overcome these difficulties and combine the switching signals and the linear modes into a single hybrid system.

This paper presents a novel approach to the study of discrete-time SLSs where the switching signals are generated by logical dynamic systems. This approach is inspired by the work in \cite{guo22} and differs from the existing literature, such as \cite{ge01,sun01,sun02,xie02}, which considers SLSs with freely chosen switching signals. The system diagram is shown in Figure \ref{fig.2}, where the top layer (supervisor or mode changer) is a logical control network with output $\sigma(t)$, generated by logical state $\gamma(t)$ and logical input $\theta(t)$, controlling the modes of the bottom layer, i.e. the discrete-time SLS.
\tikzset{
block/.style = {draw, fill=white, rectangle, minimum height=3em, minimum width=3em},
tmp/.style  = {coordinate}, 
sum/.style= {draw, fill=white, circle, node distance=1cm},
input/.style = {coordinate},
output/.style= {coordinate},
pinstyle/.style = {pin edge={to-,thin,black}
}
}

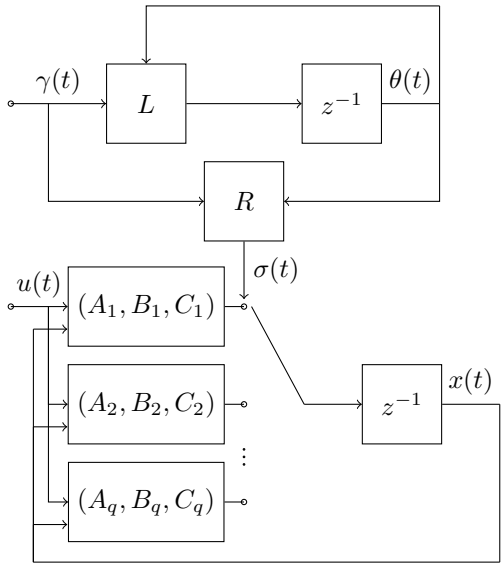
\begin{figure}[h!]
\centering
\begin{tikzpicture}[auto, node distance=1.3cm]
    \node [input, name=linput] (linput) {};
    \node [input, right of=linput, node distance=0.5cm] (sum) {};
    \node [block, right of=sum] (controller) {$L$};
    \node [input, right of=controller] (lstate) {};
    \node [block, right of=lstate] (holder) {$z^{-1}$};
    \node [block, below of=lstate] (generator) {$R$};
    \node [input, right of=holder] (loutput) {};
    \node [input, above of=loutput] (sum1) {};
    \draw [->] (linput) -- node{$\gamma(t)$} (controller);
    \draw [->] (controller) --  (holder);
    \draw [-] (holder) -- node{$\theta(t)$} (loutput);
    \draw [->] (loutput) |- (generator);
    \draw [->] (sum) |- (generator);
    \draw [-] (loutput) -- (sum1);
    \draw [->] (sum1) -| (controller);
        \node [output, below of=generator] (sum2) {};
        \draw [->] (generator) -- node{$\sigma(t)$} (sum2);
            \node [output, below of=sum2, node distance=0.1cm] (subsys1) {};
            \node [output, below of=subsys1] (subsys2) {};
            \node [output, below of=subsys2] (subsysq) {};
            \node [block, left of=subsys1] (mode1) {$(A_1,B_1,C_1)$};
            \node [block, left of=subsys2] (mode2) {$(A_2,B_2,C_2)$};
            \node [block, left of=subsysq] (modeq) {$(A_q,B_q,C_q)$};
            \node [input, right of=subsys1, node distance=0.1cm] (cross2) {};
            \node [input, right of=subsys2, node distance=0.8cm] (switch) {};
            \node [block, right of=switch] (holder1) {$z^{-1}$};
            \draw [->] (switch) --  (holder1);
            \draw [-] (mode1) --  (subsys1);
            \draw [-] (mode2) --  (subsys2);
\draw (0,0) circle (1pt);
\draw (0,-2.7) circle (1pt);
\draw (3.1,-2.7) circle (1pt);
\draw (3.1,-4) circle (1pt);
\draw (3.1,-5.3) circle (1pt);

\draw (3.1,-4.6) circle (0.3pt);
\draw (3.1,-4.7) circle (0.3pt);
\draw (3.1,-4.8) circle (0.3pt);
            \draw [-] (modeq) --  (subsysq);
            \draw [-] (switch) --  (cross2);
            \node [input, left of=mode1] (sum3) {};
            \node [input, left of=sum3, node distance=0.5cm] (input) {};
            \draw [->] (input) --  node{$u(t)$}(mode1);
            \draw [->] (sum3) |-  (mode2);
            \draw [->] (sum3) |-  (modeq);
            \node [input, right of=holder1] (output) {};
            \draw [-] (holder1) -- node{$x(t)$} (output);
            \node [input, below of=output, node distance=2.1cm] (cross) {};
            \draw [-] (output) -- (cross);
            \node [input, left of=cross, node distance=6.2cm] (cross1) {};
            \draw [-] (cross) --  (cross1); 
            \node [input, above of=cross1, node distance=0.5cm] (subq) {};
            \node [input, right of=subq, node distance=0.45cm] (subqq) {};
            \draw [-] (cross1) --  (subq); 
            \draw [->] (subq) --  (subqq); 
            \node [input, above of=subq, node distance=1.3cm] (sub2) {};
            \node [input, right of=sub2, node distance=0.45cm] (sub22) {};
            \draw [-] (cross1) --  (sub2); 
            \draw [->] (sub2) --  (sub22); 
            \node [input, above of=sub2, node distance=1.3cm] (sub1) {};
            \node [input, right of=sub1, node distance=0.45cm] (sub11) {};
            \draw [-] (cross1) --  (sub1); 
            \draw [->] (sub1) --  (sub11); 
                 \end{tikzpicture}
                 \caption{System diagram\label{fig.2}}
\end{figure}
 The proposed model is applicable to systems that contain a multi-layered structure and hybrid states, such as the cyber-physical systems (CPS) \cite{lee08,raj10,kar11,der12,lee15}. Recently, with the development of computer science, CPS has become a hot topic in the fields. A CPS has both computational and physical components and is therefore capable of performing well-defined tasks. It has been assessed by the US National Science Foundation that ``the CPSs of tomorrow will far exceed those of today in terms of adaptability, autonomy, efficiency, functionality, reliability, security, and usability''. As pointed out in the title of \cite{lee08}, the design problems of CPSs are still challenging today, and one of the several reasons is the ``lack of temporal semantics and adequate concurrency models in computing''. Another example is Game-Based Control Systems (GBCS) \cite{zha19a,zha19b,zha21}, which have been proposed to address control problems in ``social, economic, and now rapidly evolving ``intelligent'' systems. In a GBCS, there is a hierarchical decision structure in which a regulator makes his decision first, and then the agents make their strategies to pursue an optimal payoff. Understandably, the regulator makes its policy using a logical dynamic process, and the agents are modeled by linear/nonlinear dynamical systems.

 Although temporal logic still lacks a general modeling method, the logical dynamic systems of propositional logic have already been well studied with the STP approach and can be used as the computational part of a CPS. Since the birth of STP of matrices (see \cite{che11,che12}), modeling logical dynamical systems is no longer a problem. Using the algebraic state-space representation (ASSR) method based on the STP approach, a logical dynamical system can be described as a linear-like system, and one can then gain insight into the system through a structure matrix. Using the STP approach, many unsolved problems, not only in logical dynamical systems but also in finite games and other applications, have been easily solved. For example, stability and stabilisation \cite{li12,li14,men19,li20,hua21}, controllability \cite{che09a,che14,li15,che18}, observability \cite{for13,las13,zha16,guo18,zha20a,yu22,liu22}, bisimulation \cite{li18,li21,li22}, output tracking \cite{li17,zha18,zha19}, identification \cite{che11a,wan22} and many other control problems of logical dynamical systems. 

Using the ASSR method, the structures of the logical control networks and the discrete-time SLSs are unified to form a linear-like hybrid system. When analyzing the discrete-time SLSs, the advantages of the ASSR method are twofold:
\begin{itemize}
    \item The algebraic model of logical control networks in discrete-time SLSs can be embedded in the commonly used state-space model and then considered as a whole system with hybrid states.
    \item The properties of logical control networks, together with the limitations of switching signals inherited from the structure of logical control networks, do not need to be considered separately.
\end{itemize}
Based on the derived hybrid system, this paper first discusses the problems of reachability, controllability, observability, and reconstructibility of discrete-time SLSs under logic dynamic switching, and then provides criteria for the system realization problems. The corresponding necessary and sufficient conditions for the four properties are obtained, as well as an algorithm that provides an easy way to verify the control problems. 

On the other hand, when the constraints on the switching signals of an SLS take the form of dynamical systems, we are interested in whether the logical network is able to generate the desired switching signals. Such problems are crucial for evaluating the performance of switching systems. This paper provides criteria for the solvability of the realization of logical control networks with respect to two cases of constraints on the switching signals: fixed-operating-time switches, where the time of the signal to stay in each state is fixed; finite reference-signal switches, where the signals should follow a given reference signal sequence.

Using the ASSR method, in \cite{guo22}, the authors constructed logical control networks for different requirements of the generated switching sequences. Based on this, to solve the above realization problems from a different perspective, this paper provides a new set controllability method, called $\ell$-step input-state set reachability, to verify whether a logical control network can generate the desired switching signals.

The rest of this paper is organized as follows: Section 2 recalls the STP-based mergence of discrete-time SLSs. In section 3, the reachability, controllability, observability, and reonstructibility problems for discrete-time SLSs under logic dynamic switching are studied. The system realization problems based on two cases of switching signal restrictions are considered in Section 4 with the newly-introduced $\ell$-step input-space reachability method. An illustrative example is provided in Section 5. Section 6 contains the concluding remarks.

The notations used in the text are listed below:
\begin{itemize}
    \item ${\cal M}_{m\times n}$: the set of $m\times n$ real matrices.
    \item $\Col(A)$ ($\Row(A)$): the set of columns (rows) of $A$.
    \item $\Col_i(A)$ ($\Row_i(A)$): the $i$-th column (row) of $A$.
    \item ${\cal D}_k:=\{1,2,3,\cdots,k\}$.
    \item $\delta_n^i$: the $i$-th column of identity matrix $I_n$.
    \item $\Delta_n:=\Delta_n=\{\d_n^i\,|\,i=1,2,\cdots,n\}$.
    \item $[A]_{i,j}:=\Row_i(\Col_j(A))$.
    \item ${\cal L}_{m\times n}$: the set of ${m\times n}$ logical matrices, that is, $\Col({\cal L})\subset \Delta_{m}$.
    \item ${\cal B}_{m\times n}$: the set of $m\times n$ Boolean matrices, that is, $[{\cal B}]_{i,j}\in {\cal D}:=\{0,1\}$.
    \item $\d_{m}[i_1,i_2,\cdots,i_n]:=[\d_m^{i_1},\d_m^{i_2},\cdots,\d_m^{i_n}]\in {\cal L}_{m\times n}$.   
    \item ${\bf 1}_{n}:={\underbrace{[1,1,\cdots,1]}_n}^\mathrm{T}$.
  \item $\wedge,\vee,\neg$: the conjunction, disjunction, and negation operators.
  \item $A+_{\cal B}B$: the Boolean addition of $A,B\in{\cal B}_{m\times n}$ that is, $[A+_{\cal B}B]_{i,j}=[A]_{i,j}\vee [B]_{i,j}.$
  \item $A\times_{\cal B} B$: the Boolean product of $A\in{\cal B}_{m\times n},B\in{\cal B}_{n\times p}$, and $[A\times_{\cal B}B]_{i,j}=0$ if $[AB]_{i,j}=0$, $[A\times_{\cal B}B]_{i,j}=1$ if $[AB]_{i,j}>0$.
  \item $A\otimes B$: the Kronecker product of $A$ and $B$.
  \item $A*B$: the Khatri-Rao product of $A \in{\cal M}_{p\times n}$ and $B\in{\cal M}_{q\times n}$, that is $i=1,2,\cdots n$,  $\Col_i(A*B)=\Col_i(A)\otimes\Col_i(B)$.
   \item $A\wedge B:=[A\wedge B]_{i,j}=[A]_{i,j}\wedge[B]_{i,j},~A,B\in{\cal B}_{m\times n}$.
     \item  $W_{[m,n]}=\left[I_n\otimes \d_{m}^1,I_n\otimes \d_{m}^2,\ldots,I_n\otimes \d_{m}^m \right]$: the swap matrix, that is, for $x\in\Delta_m,y\in\Delta_n$, $W_{[m,n]}x\otimes y=y\otimes x$.
      \item $\Phi_n:=\delta_{n^2}[1,n+2,2n+3,\cdots,(n-2)n+(n-1),n^2]$ is the power-reducing matrix. It follows that $\forall x\in\Delta_n$, $\Phi_n x=x^2$.
\item $|S|$: the cardinality of a set $S$.
\item $[a,b]:=\{a,a+1,\cdots,b\}$, where $a,b\in\mathbb{Z},a<b$.
\item $\mathrm{Im(A)}$: the image of a matrix $A$.
\end{itemize}

The main mathematical tool used in this paper is the STP of matrices, denoted by ``$\ltimes$'', which is omitted if there is no confusion. For details, please refer to Appendix. 

\section{STP-Based Mergence}

We consider linear discrete-time switched systems where the switching signals are generated by logic dynamical systems. The following equations depict the model:
\begin{align}
\label{eq3.1}\left\{
\begin{array}{lll}
x(t+1)=A_{\sigma(t)}x(t)+B_{\sigma(t)}u(t),\\
y(t)=C_{\sigma(t)}x(t),
\end{array}\right.
\end{align}
\begin{align}\label{eq3.2.0}
\begin{cases}
\theta_1(t+1)=f_1(\theta_1(t),\cdots,\theta_k(t);\gamma_{_1}(t),\cdots,\gamma_{l}(t)),\\
\theta_2(t+1)=f_2(\theta_1(t),\cdots,\theta_k(t);\gamma_{_1}(t),\cdots,\gamma_{l}(t)),\\
~\quad\quad\quad\quad\vdots\\
\theta_k(t+1)=f_k(\theta_1(t),\cdots,\theta_k(t);\gamma_{_1}(t),\cdots,\gamma_{l}(t)),\\
\sigma(t)=h(\theta_1(t),\cdots,\theta_k(t);\gamma_{_1}(t),\cdots,\gamma_{l}(t)),
\end{cases}
\end{align}
where $x(t)\in\mathbb{R}^n, u(t)\in\mathbb{R}^m, y(t)\in\mathbb{R}^p$ are the state, input, and output, respectively; $\sigma(t)\in[1,q]$ is the switching signal; $(A_i,B_i,C_i),~i\in[1,q]$ are the subsystems, also known as the modes, of the discrete-time SLS (\ref{eq3.1}); (\ref{eq3.2.0}) is a (mixed-value) logcial control network with ${\theta}_i(t)\in{\cal{D}}_{N_i},~i\in[1, k]$ being the logical states, ${\gamma}_j(t)\in{\cal{D}}_{M_j}~j\in[1, l]$ being the logical inputs and ${\sigma}(t)\in{\cal{D}}_{q}$ being the switching signal. $f_i:\prod\limits_{i=1}^k{\cal D}_{N_i}\times\prod\limits_{j=1}^l{\cal D}_{M_j}\rightarrow {\cal D}_{N_i}$, $i\in[1,k],j\in[1,l]$, and $h:\prod\limits_{i=1}^k{\cal D}_{N_i}\times\prod\limits_{j=1}^l{\cal D}_{M_j}\rightarrow {\cal D}_{q}$, $i\in[1,k],j\in[1,l]$ are logical functions.

The ASSR of the logical control network (\ref{eq3.2.0}) can be obtained by resorting to the STP approach (please refer to Appendix for the details on the transformation process between the algebraic and vector forms of the logical variables) as follows:
\begin{align}\label{eq3.2}\left\{
\begin{array}{lll}
\vec{\theta}(t+1)=L\ltimes \vec{\gamma}(t)\ltimes \vec{\theta}(t),\\
\vec{\sigma}(t)=R\ltimes \vec{\gamma}(t)\ltimes \vec{\theta}(t),
\end{array}\right.
\end{align}
 where $\vec{\theta}(t)\in\Delta_{N},~N:=\prod\limits_{i=1}^kN_i$, $\vec{\gamma}(t)\in\Delta_M,~M:=\prod\limits_{j=1}^lN_j$, $\vec{\sigma}(t)\in\Delta_{q}$ are the overall logical state, overall logical input, and switching signal in vector forms; $L\in{\cal L}_{N\times MN}$ and $R\in{\cal L}_{q\times MN}$ are the structure matrix and signal-generating matrix of (\ref{eq3.2}). $L$ is often written as $[L_1,L_2,\cdots,L_M]$, where $L_i\in{\cal L}_{N\times N}$.

Using the STP approach, equation (\ref{eq3.1}) can be rewritten as
\begin{align}
\label{eq3.3}
\left\{
\begin{array}{lll}
x(t+1)={\bf A}\ltimes \vec{\sigma}(t)\ltimes x(t)+{\bf B}\ltimes \vec{\sigma}(t)\ltimes u(t),\\
y(t)={\bf C}\ltimes \vec{\sigma}(t)\ltimes x(t),
\end{array}\right.
\end{align}
where ${\bf A}:=[A_1~A_2~\cdots~A_q],~{\bf B}:=[B_1~B_2~\cdots~B_q]$, and ${\bf C}:=[C_1~C_2~\cdots~C_q]$.
 
Letting $z(t):=\vec{\theta}(t) \ltimes x(t)\in 
\mathbb{R}^{Nn}$, the STP-based mergence of systems (\ref{eq3.3}) and (\ref{eq3.2}) is derived as follows \cite{guo22}:
 \begin{align}\label{eq3.4}
 \begin{array}{lll}
 z(t+1)&=&\vec{\theta}(t+1)\ltimes x(t+1)\\
&=& L \vec{\gamma}(t) \vec{\theta}(t)[{\bf A} \vec{\sigma}(t)  x(t)+{\bf B}\vec{\sigma}(t) u(t)]\\
&=&L \vec{\gamma}(t) \vec{\theta}(t){\bf A}R\vec{\gamma}(t) \vec{\theta}(t)x(t)\\&~&+L \vec{\gamma}(t) \vec{\theta}(t){\bf B}R\vec{\gamma}(t) \vec{\theta}(t)u(t)\\
&=& L[I_{MN}\otimes ({\bf A}R)]\Phi_{MN}\vec{\gamma}(t) \vec{\theta}(t)x(t)\\
&~&+L[I_{MN}\otimes ({\bf B}R)]\Phi_{MN}\vec{\gamma}(t) \vec{\theta}(t)u(t)\\
&=& {\bf G}\vec{\gamma}(t) z(t)+{\bf H}\vec{\gamma}(t) \vec{\theta}(t)u(t),
\end{array}
\end{align}
where ${\bf G}:=L[I_{MN}\otimes ({\bf A}R)]\Phi_{MN}\in {\cal M}_{nN\times nMN}$, ${\bf H}:=L[I_{MN}\otimes ({\bf B}R)]\Phi_{MN}\in {\cal M}_{nN\times mMN}$.

Since $\vec{\gamma}(t)\in \Delta_{M}$, based on the properties of the STP, the merged system (\ref{eq3.4}) can be rewritten as:
\begin{align}\label{eq3.5}
\begin{array}{lll}
z(t+1)&=&[G_1~G_2~\cdots~G_M]\vec{\gamma}(t) z(t)\\&~&+[H_1~H_2~\cdots~H_M]\vec{\gamma}(t) \vec{\theta}(t)u(t),
\end{array}
\end{align}
where $$G_i=\begin{bmatrix} G^i_{1,1}&G^i_{1,2}&\cdots&G^i_{1,N}\\
G^i_{2,1}&G^i_{2,2}&\cdots&G^i_{2,N}\\
\vdots&\vdots&~&\vdots\\
G^i_{N,1}&G^i_{N,2}&\cdots&G^i_{N,N}
\end{bmatrix},$$ and $$H_i=\begin{bmatrix} H^i_{1,1}&H^i_{1,2}&\cdots&H^i_{1,N}\\
H^i_{2,1}&H^i_{2,2}&\cdots&H^i_{2,N}\\
\vdots&\vdots&~&\vdots\\
H^i_{N,1}&H^i_{N,2}&\cdots&H^i_{N,N}
\end{bmatrix},$$
each submatrix $G^i_{\alpha,\beta}\in {\cal M}_{n\times n}$ and $H^i_{\alpha,\beta}\in {\cal M}_{n\times m}$, $\alpha,\beta\in[1,N]$. We call the above matrices the block form of $G_i$ and $H_i$, $i=1,\cdots,M$.



Using the merged system (\ref{eq3.5}), we formally eliminated the constraint that the switching signals $\sigma(t)$ must be produced by the logical control network (\ref{eq3.2}). One can see that the logical inputs $\gamma(t)$, which can be selected arbitrarily, are the ``switching signals" of the system (\ref{eq3.5}), and the original switching signals $\sigma(t)$ are implicit at present. 

\begin{rem}
    \label{r2.1}
    In the subsequent sections, the logical state and input variables $\theta$ and $\gamma$ appear as the overall variables unless otherwise specified. For example, $(\gamma_1,\gamma_2,\cdots,\gamma_T)$ represents a logical input sequence but not a series of logical input nodes.
\end{rem}
\section{Main Results}

\subsection{Reachability Problem}
\begin{dfn}\label{d4.1}
Consider the switched linear system (\ref{eq3.1}) in which the switching signal is generated by the logical control network (\ref{eq3.2}).
\begin{enumerate}
  \item A state $x\in \mathbb{R}^n$ is reachable, if there exists a positive integer $T<\infty$, a logical input sequence $(\gamma_0,\gamma_1,\cdots,\gamma_{T-1})$, and an input sequence $(u_0,u_1,\cdots,u_{T-1})$, such that for {$\forall \vec{\theta}_0\in\Delta_N$},
  $$x_0={\bf 0}, \quad x_T=x.$$
  \item The system is reachable if all states in $\mathbb{R}^n$ are reachable.
  \end{enumerate}
\end{dfn}

 \begin{dfn}
 \label{d4.2}
Consider the merged system (\ref{eq3.5}). Define the reachable set of state $x={\bf 0}$ with $T$-length logical input sequence $(\gamma_0,\gamma_1,\cdots,\gamma_{T-1})$ and initial logical state $\alpha\in [1,N]$ as
\begin{align}\label{eq5.0}
\begin{array}{c}
 {\cal R}_T^\alpha(\gamma_0,\gamma_1\cdots,\gamma_{T-1}):=\\
 {\bf 1}_N^\mathrm{T}\left[\mathrm{Im}(G_{\gamma_{T-1}}G_{\gamma_{T-2}}\cdots G_{\gamma_{1}}H_{\gamma_0}\delta_N^\alpha)\right]\cup \\
{\bf 1}_N^\mathrm{T}\left[\mathrm{Im}(G_{\gamma_{T-1}}G_{\gamma_{T-2}}\cdots G_{\gamma_{2}}H_{\gamma_1}L_{\gamma_0}\delta_N^\alpha)\right]\cup\\
  \vdots \\\cup  {\bf 1}_N^\mathrm{T}\left[\mathrm{Im}(G_{\gamma_{T-1}}H_{\gamma_{T-2}}L_{\gamma_{T-3}}\cdots L_{\gamma_1}L_{\gamma_0}\delta_N^\alpha)\right]\\
\cup {\bf 1}_N^\mathrm{T}\left[\mathrm{Im}(H_{\gamma_{T-1}}L_{\gamma_{T-2}}\cdots L_{\gamma_0}\delta_N^\alpha)\right].
\end{array}
\end{align}
 \end{dfn}
Now we are ready for the following result.
\begin{thm}\label{t4.1}
The switched linear system (\ref{eq3.1}) is reachable under the logically generated switching signal (\ref{eq3.2}), if and only if there exists a logical input sequence $(\gamma_0,\gamma_1,\cdots,\gamma_{T-1})$, such that the reachable set of the merged system (\ref{eq3.5}) satisfies
\begin{align}\label{eq4.1}
\bigcap\limits_{\alpha=1}^N {\cal R}_T^\alpha(\gamma_0,\gamma_1\cdots,\gamma_{T-1})=\mathbb{R}^n.
\end{align}
\end{thm}

{\it Proof:}
Before proving Theorem \ref{t4.1}, we first have a closer look into the structures of ${\bf G}$ and ${\bf H}$.

From equation (\ref{eq3.4}), one has
\begin{align*}
\begin{array}{lll}
{\bf G}&=&L\begin{bmatrix}
{\bf A}R&~&~\\
~&\ddots &~\\
~&~&{\bf A}R 
\end{bmatrix}
\begin{bmatrix}\delta_{(MN)^2}^1&~&~\\
~&\ddots&~\\
~&~&\delta_{(MN)^2}^{(MN)^2}\end{bmatrix}\\
&=&L\begin{bmatrix}
{\bf A}(R\delta_{MN}^1)\delta_{MN}^1&~&~\\
~&\ddots&~\\
~&~&{\bf A}(R\delta_{MN}^{MN})\delta_{MN}^{MN}
\end{bmatrix}\\
&=&L\begin{bmatrix}
A_{\sigma^1}\delta_{MN}^1&~&~\\
~&\ddots&~\\
~&~&A_{\sigma^{MN}}\delta_{MN}^{MN}
\end{bmatrix}\\
&=&\begin{bmatrix} \Col_1(L)A_{\sigma^1}&\cdots&\Col_{MN}(L)A_{\sigma^{MN}}    \end{bmatrix},
\end{array}
\end{align*}
where $A_{\sigma^i}:=A_{\sigma}|_{\vec{\sigma}=\Col_i(R)}$.

Then it is straightforward that
{\scriptsize
\begin{align*}
\begin{array}{c}
G_1=\begin{bmatrix}\Col_1(L)A_{\sigma^1}&\cdots&\Col_{N}(L)A_{\sigma^N}\end{bmatrix},\\
G_2=\begin{bmatrix}\Col_{N+1}(L)A_{\sigma^{N+1}}&\cdots&\Col_{2N}(L)A_{\sigma^{2N}}\end{bmatrix},\\
\vdots\\
G_M=\begin{bmatrix}\Col_{(M-1)N+1}(L)A_{\sigma^{(M-1)N+1}}&\cdots&\Col_{MN}(L)A_{\sigma^{MN}}\end{bmatrix}.\\
\end{array}
\end{align*}}
Since $L$ is a logical matrix, $\Col(L)\subset \Delta_{N}$, in the block form of $G_i$ there is only one nonzero block in each column. Furthermore, if we compress each nonzero block into $1$, zero block into $0$, and denote the compressed matrix by $\tilde{G}_i$, then $\tilde{G}_i=L_i$ and consequently, $\tilde{{\bf G}}=L$. It is obvious that ${\bf H}$ also has the above properties. Now we are ready for the proof.

{From the structure of ${\bf G}$ and ${\bf H}$, we know that $G_{\gamma_{T-1}}G_{\gamma_{T-2}}\cdots G_{\gamma_{1}}H_{\gamma_0}$ also has the block form where each column has only one nonzero block, so do the other matrices in similar forms on the right side of equation (\ref{eq5.0}).}

We first analyse the term $${\bf 1}_N^\mathrm{T}\left[\mathrm{Im}(G_{\gamma_{T-1}}G_{\gamma_{T-2}}\cdots G_{\gamma_{1}}H_{\gamma_0}\delta_N^\alpha)\right]$$ in equation (\ref{eq5.0}). In the $\alpha$-th column of the block form of $G_{\gamma_{T-1}}G_{\gamma_{T-2}}\cdots G_{\gamma_{1}}H_{\gamma_0}$ (i.e. $G_{\gamma_{T-1}}G_{\gamma_{T-2}}\cdots G_{\gamma_{1}}H_{\gamma_0}\delta_N^\alpha$, $\alpha\in[1,N]$), the nonzero block has the following form: 
$$G^{\gamma_{T-1}}_{\upsilon,\epsilon}G^{\gamma_{T-2}}_{\epsilon,\xi}\cdots G^{\gamma_{1}}_{\kappa,\beta}H^{\gamma_0}_{\beta,\alpha}=A_{\sigma_{T-1}}A_{\sigma_{T-2}}\cdots A_{\sigma_1}B_{\sigma_0},$$
where $(\gamma_0,\gamma_1,\cdots,\gamma_{T-1})$ is the logical input sequence, $(\alpha,\beta,\kappa,\cdots,\xi,\epsilon,\upsilon)$ and $(\sigma_{0},\sigma_{1},\cdots,\sigma_{T-1})$ are the corresponding logical state trajectory and switching signal trajectory, respectively (one should notice that the length of the logical trajectory is $T+1$). To be precise, $\sigma_{T-1}=\sigma^{N(\gamma_{T-1}-1)+\epsilon},$ $\sigma_{T-2}=\sigma^{N(\gamma_{T-2}-1)+\xi},\cdots,\sigma_0=\sigma^{N(\gamma_{0}-1)+\alpha}$. Thus, 
\begin{align*}
\begin{array}{c}
\mathrm{Im}(G_{\gamma_{T-1}}G_{\gamma_{T-2}}\cdots G_{\gamma_{1}}H_{\gamma_0}\delta_N^\alpha)\\=\mathrm{Im}(\delta_N^{\upsilon} A_{\sigma_T}A_{\sigma_{T-1}}\cdots A_{\sigma_1}B_{\sigma_0})\subset \{\delta_N^{\upsilon}\}\times \mathbb{R}^n.
\end{array}
\end{align*}
It is obvious that
\begin{align*}
\begin{array}{c}
{\bf 1}_N^\mathrm{T}\left[\mathrm{Im}(G_{\gamma_{T-1}}G_{\gamma_{T-2}}\cdots G_{\gamma_{1}}H_{\gamma_0}\delta_N^\alpha)\right]\\=\mathrm{Im}(A_{\sigma_{T-1}}A_{\sigma_{T-2}}\cdots A_{\sigma_1}B_{\sigma_0}).
\end{array}
\end{align*}

 The other terms in equation (\ref{eq5.0}) have similar results. Then it means that with the initial logical state $\alpha$, there is a switching signal trajectory $(\sigma_0,\sigma_1,\cdots,\sigma_{T-1})$, such that
\begin{align*}
\begin{array}{ccc}
\mathrm{Im}(A_{\sigma_{T-1}}A_{\sigma_{T-2}}\cdots A_{\sigma_1}B_{\sigma_0})\cup\\
\mathrm{Im}(A_{\sigma_{T-1}}A_{\sigma_{T-2}}\cdots A_{\sigma_2}B_{\sigma_1})\cup\\
 \cdots \\
 \cup \mathrm{Im}(A_{\sigma_{T-1}}B_{\sigma_{T-2}})\cup \mathrm{Im}(B_{\sigma_{T-1}})=\mathbb{R}^n,
\end{array}
 \end{align*}
which means that starting from $x(0)={\bf 0}$ and $\vec{\theta}_0=\delta_N^\alpha$, with certain input sequence $(u_0,u_1,\cdots,u_{T-1})$, $$x(T)=A_{\sigma_{T-1}}A_{\sigma_{T-2}}\cdots A_{\sigma_1}B_{\sigma_0}u_0+\cdots+B_{\sigma_{T-1}}u_{T-1}$$ can reach anywhere in $\mathbb{R}^n$.

The above analysis is based on the initial logical state $\alpha$. To eliminate this constraint, $\alpha$ should be any possible value. Otherwise, the logical input sequence $(\gamma_0,\gamma_1\cdots,\gamma_{T-1})$ is not feasible for some initial logical states such that the vector space $\mathbb{R}^n$ cannot be spanned by the corresponding switching signal sequences. Now the proof is completed.
\hfill $\Box$

\begin{rem}\label{r4.1}
In Theorem \ref{t4.1}, one may not have to take all $\alpha\in[1,N]$ into the computation. If there is a $\beta\in[1, N]$ such that equation (\ref{eq4.1}) holds, then one does not have to check the states that can be controlled to $\beta$, because, for these states, one can first drive them to $\beta$ and then solve the problem. To be precise, the number of the to-be-checked states is the number of the disjoint control attractors (i.e., the attractors whose attract basins are disjoint) of the logical system (\ref{eq3.2}).
\end{rem}

    It should be pointed out that Theorem \ref{t4.1} in its present form can hardly serve as an easily verifiable criterion for controllability, because one may find it difficult to check the existence of the logical input sequences required by (\ref{d4.1}). Thus, in the following, we provide an algorithm to overcome such difficulty.

By the proof of Theorem \ref{t4.1} and Remark \ref{r4.1}, the following algorithm gives the feasible logical input sequences.

\begin{alg}\label{a4.3} Computing the feasible logical input sequences.
\begin{enumerate}
  \item [Step]1: Set $k:=1$, go to Step 2.
  \item [Step]2: Compute 
\begin{align*}
  {\cal G}_{k}:=\left( \sum\limits_{i=1}^M G_i \right)^{k},
  \end{align*}
  then go to Step 3.\\
  \%here is the symbolic computation on the block form
  \item [Step]3: Use Proposition 7 in \cite{zha19} to find all the control attractors in the whole logical state space, and for the attractors whose attract basins contain some common logical states, just pick one of them for the following computation.\\
   For a control attractor $\alpha$, each nonzero block on the $\alpha$-th column of ${\cal G}_{k}$ has the following form (assume that the $(\upsilon,\alpha)$-th block of ${\cal G}(k)$ is nonzero): 
\begin{align*}
\begin{array}{lll}
  G^{\gamma^j_{k-1}}_{\upsilon,\epsilon^j}G^{\gamma^j_{k-2}}_{\epsilon^j,\xi^j}\cdots G^{\gamma^j_0}_{\beta^j,\alpha}+G^{\gamma^\ell_{k-1}}_{\upsilon,\epsilon^\ell}G^{\gamma^\ell_{k-2}}_{\epsilon^\ell,\xi^\ell}\cdots G^{\gamma^\ell_0}_{\beta^\ell,\alpha}+\cdots .
  \end{array}
  \end{align*}
  Go to Step 4.
  \item [Step]4: Take each alternative logical input sequence $(\gamma^j_0,\cdots,\gamma^j_{k-2},\gamma^j_{k-1})$ and its corresponding logical state trajectory $(\alpha,\beta^j,\cdots,\xi^j,\epsilon^j,\upsilon)$ into equation (\ref{eq4.1}).\\
  If (\ref{eq4.1}) holds for all the initial logical states which are control attractors (see Definition \ref{d3.1} of Appendix, and only one logical state in a control cycle needs to be checked), then $(\gamma^j_0,\cdots,\gamma^j_{k-2},\gamma^j_{k-1})$ is the desirable logical input sequence, stop.\\
  Otherwise, set $k:=k+1$, and go to Step 2.
\end{enumerate}
\end{alg}

\begin{rem}
    \label{r4.3}
    In Algorithm \ref{a4.3}, an alternative logical input sequence $(\gamma_0,\gamma_1,\cdots,\gamma_{T-1})$ only needs to be checked for a few control attractors whose attract basins compose the whole state space. Because for the other states, the corresponding logical input sequences can be obtained by adding the segments $(\gamma_{-\tau},\gamma_{-\tau+1},\cdots,\gamma_{-1})$, which can drive them to the control attractors, to the left side of the checked logical input sequence.
\end{rem}

\begin{rem}
\label{r4.4}
From the proof of Theorem \ref{t4.1}, we can see that the criterion (\ref{eq4.1}) for the reachability of the merged system (\ref{eq3.5}) is equivalent to the Kalman-type rank condition for the reachability of the system (\ref{eq3.1}), that is, the system is reachable, if and only if there exists a switching signal sequence $(\sigma_0,\sigma_1,\cdots,\sigma_{T-1})$ generated by the system (\ref{eq3.2}), such that
$$\mathrm{rank}[B_{\sigma_{T-1}},A_{\sigma_{T-1}}B_{\sigma_{T-2}},\cdots,A_{\sigma_{T-1}}\cdots A_{\sigma_{1}}B_{\sigma_{0}}]=n.$$ But one should notice that without the proposed approach, the Kalman-type rank condition cannot be applied to the studied hybrid systems because the system (\ref{eq3.5}) which contains both physical and logical states, is not a linear system. Besides, one cannot guarantee whether the switching signal sequence $(\sigma_0,\sigma_1,\sigma_{T-1})$ can be generated by the logical control network. Hence, the Kalman-type rank condition is not feasible for solving the problems considered in this paper.
\end{rem}

\subsection{Controllability Problem}
\begin{dfn}\label{d4.3}
Consider the switched linear system (\ref{eq3.1}) in which the switching signal is generated by the logical control network (\ref{eq3.2}).
\begin{enumerate}
  \item A state $x\in \mathbb{R}^n$ is controllable, if there exist a positive integer $T<\infty$, a logical input sequence $(\gamma_0,\gamma_1,\cdots,\gamma_{T-1})$, and an input sequence $(u_0,u_1,\cdots,u_{T-1})$, such that for $\forall\vec{\theta}_0\in\Delta_{N}$, $$x_0=x,\quad x_T={\bf 0}.$$
  \item The system is controllable if all the states in $\mathbb{R}^n$ are controllable.
  \end{enumerate}
\end{dfn}

\begin{thm}\label{t4.2}
The switched linear system (\ref{eq3.1}) is controllable under the logically generated switching signal, if and only if there exists a logical input sequence $(\gamma_0,\gamma_1,\cdots,\gamma_{T-1}),~T<\infty$, such that starting from any logical state $\alpha\in[1,N]$, the merged system (\ref{eq3.5}) satisfies
\begin{align}\label{eq4.2}
 {\bf 1}_N^\mathrm{T}\left[\mathrm{Im}(G_{\gamma_{T-1}}G_{\gamma_{T-2}}\cdots G_{\gamma_0}\delta_N^\alpha)\right]\subset{\cal R}_T^\alpha(\gamma_0,\gamma_1\cdots,\gamma_{T-1}).
\end{align}
\end{thm}

{\it Proof:} By equation (\ref{eq3.5}), if the system is controllable, then there exists $T<\infty$ such that the following equation holds:
\begin{align*}
\begin{array}{c}
    z(T)=G_{\gamma_{T-1}}\cdots G_{\gamma_0}z(0)+G_{\gamma_{T-1}}\cdots G_{\gamma_{1}}H_{\gamma_0}\vec{\theta}(0)u(0)\\
   +G_{\gamma_{T-1}}\cdots G_{\gamma_{2}}H_{\gamma_1}\vec{\theta}(1)u(1)+
    \cdots
    \\+G_{\gamma_{T-1}}H_{\gamma_{T-2}}\vec{\theta}(T-2)u(T-2)\\
    +H_{\gamma_{T-1}}\vec{\theta}(T-1)u(T-1)={\bf 0}.
\end{array}
\end{align*}
Then for some initial logical state denoted by $\delta^\alpha_N$, one has
\begin{align*}
\begin{array}{r}
    G_{\gamma_{T-1}}\cdots G_{\gamma_0}\delta_N^\alpha x(0)=-(G_{\gamma_{T-1}}\cdots H_{\gamma_0}\delta_N^\alpha u(0)+\cdots\\
    +G_{\gamma_{T-1}}H_{\gamma_{T-2}}L_{\gamma_{T-3}}\cdots L_{\gamma_0}\delta_N^\alpha u(T-2)\\+H_{\gamma_{T-1}}L_{\gamma_{T-2}}\cdots L_{\gamma_0}\delta_N^\alpha u(T-1)).
\end{array}
\end{align*}
According to the proof of Theorem \ref{t4.1}, the term $$G_{\gamma_{T-1}}\cdots G_{\gamma_0}\delta_N^\alpha$$ has only one nonzero block. Without loss of generality, we assume that it is the $(\upsilon,\alpha)$-th block: $$G^{\gamma_{T-1}}_{\upsilon,\epsilon}G^{\gamma_{T-2}}_{\epsilon,\xi}\cdots G^{\gamma_{1}}_{\kappa,\beta}G^{\gamma_0}_{\beta,\alpha}.$$ Then starting with logical state $\alpha$, the controllability is guaranteed if and only if for $\forall x_0\in \mathbb{R}^n$, $$(G^{\gamma_{T-1}}_{\upsilon,\epsilon}G^{\gamma_{T-2}}_{\epsilon,\xi}\cdots G^{\gamma_{1}}_{\kappa,\beta}G^{\gamma_0}_{\beta,\alpha})x_0\in {\cal R}^\alpha(\gamma_0,\gamma_1\cdots,\gamma_{T-1}),$$
which coincides with (\ref{eq4.2}). Similar with the argument in the proof of Theorem \ref{t4.1}, equation (\ref{eq4.2}) should be valid on $\forall\alpha\in[1,N]$.
\hfill $\Box$

\begin{rem}
\label{r4.2}
\begin{enumerate}
    \item When checking if equation (\ref{eq4.2}) holds for $\forall\alpha\in[1,N]$, one can also simplify the computation using Remark \ref{r4.1}.
    \item It is obvious that when the system (\ref{eq3.1}) is reversible, i.e., $A_i$ is non-singular for $\forall i\in[1,q]$, Theorems \ref{t4.1} and \ref{t4.2} are equivalent.
\end{enumerate}
\end{rem}

\begin{rem}
\label{r4.5}
We may borrow the result in \cite{ge01} to obtain the upper bound of $T$ in Theorems \ref{t4.1} and \ref{t4.2}, that is, $T\leq n$.
This upper bound is also feasible for the results in the subsequent subsection.
\end{rem}

\subsection{Observability and Reconstructibility Problems}
\begin{dfn}
\label{d5.1}
Consider the switched linear system (\ref{eq3.1}) where the switching signal is generated by the logical control network (\ref{eq3.2}). The system is observable (reconstructible), if there exists a positive integer $T<\infty$ and a logical input sequence $(\gamma_0,\gamma_1,\cdots,\gamma_{T-1})$, such that the input sequence $(u_0,u_1,\cdots,u_{T-1})$ and output trajectory $(y_0,y_1,\cdots,y_{T})$ can uniquely determine the initial state $x_0$ (the current state $x_{T}$), regardless of the value of the initial logical state $\theta_0$.
\end{dfn}


Now we build a dual system as
\begin{align}
\label{eq5.3}
\tilde{x}(t+1)={\bf \tilde{A}}\ltimes \vec{\sigma}(t)\ltimes \tilde{x}(t)+{\bf \tilde{C}}\ltimes \vec{\sigma}(t)\ltimes \tilde{u}(t),
\end{align}
where $\tilde{x}\in\mathbb{R}^n$, $\tilde{u}\in \mathbb{R}^p$, ${\bf \tilde{A}}=[\tilde{A}_1~\tilde{A}_2~\cdots~\tilde{A}_q]:=[A_1^\mathrm{T}~A_2^\mathrm{T}~\cdots~A_q^\mathrm{T}]$, ${\bf \tilde{C}}=[\tilde{C}_1~\tilde{C}_2~\cdots~\tilde{C}_q]:=[C_1^\mathrm{T}~C_2^\mathrm{T}~\cdots~C_q^\mathrm{T}]$.

The logical control network remains the same as (\ref{eq3.2}).

Similar with the construction of system (\ref{eq3.4}), we build a merged system using systems (\ref{eq5.3}) and (\ref{eq3.2}):
 \begin{align}\label{eq5.5}
 \begin{array}{lll}
 \tilde{z}(t+1)&=&\vec{\theta}(t+1)\ltimes \tilde{x}(t+1)\\
&=& L \vec{\gamma}(t) \vec{\theta}(t)[{\bf \tilde{A}} \vec{\sigma}(t)  \tilde{x}(t)+{\bf \tilde{C}}\vec{\sigma}(t) \tilde{u}(t)]\\
&=&L \vec{\gamma}(t) \vec{\theta}(t){\bf \tilde{A}}R\vec{\gamma}(t) \vec{\theta}(t)\tilde{x}(t)\\&~&+L \vec{\gamma}(t) \vec{\theta}(t){\bf \tilde{C}}R\vec{\gamma}(t) \vec{\theta}(t)\tilde{u}(t)\\
&=& L[I_{MN}\otimes ({\bf \tilde{A}}R)]\Phi_{MN}\vec{\gamma}(t) \vec{\theta}(t)\tilde{x}(t)\\
&~&+L[I_{MN}\otimes ({\bf \tilde{C}}R)]\Phi_{MN}\vec{\gamma}(t) \vec{\theta}(t)\tilde{u}(t)\\
&=& {\bf \tilde{G}}\vec{\gamma}(t) \tilde{z}(t)+{\bf \tilde{H}}\vec{\gamma}(t) \vec{\theta}(t)\tilde{u}(t),
\end{array}
\end{align}
where ${\bf \tilde{G}}:=L[I_{MN}\otimes ({\bf \tilde{A}}R)]\Phi_{MN}\in {\cal M}_{nN\times nMN}$, ${\bf \tilde{H}}:=L[I_{MN}\otimes ({\bf \tilde{C}}R)]\Phi_{MN}\in {\cal M}_{nN\times mMN}$.

System (\ref{eq5.5}) can also be rewritten as
\begin{align}\label{eq5.6}
\begin{array}{lll}
\tilde{z}(t+1)&=&[\tilde{G}_1~\tilde{G}_2~\cdots~\tilde{G}_M]\vec{\gamma}(t) \tilde{z}(t)\\&~&+[\tilde{H}_1~\tilde{H}_2~\cdots~\tilde{H}_M]\vec{\gamma}(t) \vec{\theta}(t)\tilde{u}(t),
\end{array}
\end{align}
where $$\tilde{G}_i=\begin{bmatrix} \tilde{G}^i_{1,1}&\tilde{G}^i_{1,2}&\cdots&\tilde{G}^i_{1,N}\\
\tilde{G}^i_{2,1}&\tilde{G}^i_{2,2}&\cdots&\tilde{G}^i_{2,N}\\
\vdots&\vdots&~&\vdots\\
\tilde{G}^i_{N,1}&\tilde{G}^i_{N,2}&\cdots&\tilde{G}^i_{N,N}
\end{bmatrix},$$ and $$\tilde{H}_i=\begin{bmatrix} \tilde{H}^i_{1,1}&\tilde{H}^i_{1,2}&\cdots&\tilde{H}^i_{1,N}\\
\tilde{H}^i_{2,1}&\tilde{H}^i_{2,2}&\cdots&\tilde{H}^i_{2,N}\\
\vdots&\vdots&~&\vdots\\
\tilde{H}^i_{N,1}&\tilde{H}^i_{N,2}&\cdots&\tilde{H}^i_{N,N}
\end{bmatrix},$$
with each submatrix $\tilde{G}^i_{\alpha,\beta}\in {\cal M}_{n\times n}$ and $\tilde{H}^i_{\alpha,\beta}\in {\cal M}_{n\times p}$, $\alpha,\beta\in[1,N]$. We call the above matrices the block form of $\tilde{G}_i$ and $\tilde{H}_i$, $i=1,\cdots,M$.

\begin{dfn}
\label{d5.3}
Consider the merged system (\ref{eq5.5}). Define the reachable set of state $\tilde{x}$ with $T$-length logical input sequence $(\gamma_0,\gamma_1,\cdots,\gamma_{T-1})$ and initial logical state $\alpha\in [1,N]$ as
\begin{align}\label{eq4.0}
\begin{array}{c}
 \tilde{\cal R}_T^\alpha(\gamma_{0},\gamma_{1}\cdots,\gamma_{T-1}):=\\
 {\bf 1}_N^\mathrm{T}\left[\mathrm{Im}(\tilde{G}_{\gamma_{0}}\tilde{G}_{\gamma_{1}}\cdots \tilde{G}_{\gamma_{T-2}}\tilde{H}_{\gamma_{T-1}}L_{\gamma_{T-2}}\cdots L_{\gamma_{0}}\delta_N^\alpha)\right]\cup \\
{\bf 1}_N^\mathrm{T}\left[\mathrm{Im}(\tilde{G}_{\gamma_{0}}\tilde{G}_{\gamma_{1}}\cdots \tilde{G}_{\gamma_{T-3}}\tilde{H}_{\gamma_{T-2}}L_{\gamma_{T-3}}\cdots L_{\gamma_{0}}\delta_N^\alpha)\right]\cup\\
  \vdots \\\cup  {\bf 1}_N^\mathrm{T}\left[\mathrm{Im}(\tilde{G}_{\gamma_{0}}\tilde{H}_{\gamma_{1}}L_{\gamma_{0}}\delta_N^\alpha)\right]\\
\cup {\bf 1}_N^\mathrm{T}\left[\mathrm{Im}(\tilde{H}_{\gamma_{0}}\delta_N^\alpha)\right].
\end{array}
\end{align}
 \end{dfn}

By the duality principle, the following theorem is obtained.
\begin{thm}\label{t5.1}
The switched linear system (\ref{eq3.1}) is observable under the logically generated switching signal, if and only if there exists a logical input sequence $(\gamma_0,\gamma_1,\cdots,\gamma_{T-1})$, such that the reachable set of the merged system (\ref{eq5.5}) satisfies
\begin{align}\label{eq5.7}
\bigcap\limits_{\alpha=1}^N \tilde{{\cal R}}_T^\alpha(\gamma_0,\gamma_1\cdots,\gamma_{T-1})=\mathbb{R}^n.
\end{align}
\end{thm}

{\it Proof:} Using the proof of Theorem \ref{t4.1}, one can conclude that criterion (\ref{eq5.7}) is equivalent to that for the system (\ref{eq5.3}), there exists a switching signal sequence $(\sigma_0,\sigma_1,\cdots,\sigma_{T-1})$, which is generated by the system (\ref{eq3.2}) with initial state $\alpha$, such that
\begin{align*}
\begin{array}{cc}
\mathrm{Im}\left(\tilde{A}_{\sigma_{0}}\tilde{A}_{\sigma_{1}}\cdots \tilde{A}_{\sigma_{T-2}}\tilde{C}_{\sigma_{T-1}}\right)\cup\\
\mathrm{Im}\left(\tilde{A}_{\sigma_{0}}\tilde{A}_{\sigma_{1}}\cdots \tilde{A}_{\sigma_{T-3}}\tilde{C}_{\sigma_{T-2}}\right)\cup\\
\vdots\\
\cup\mathrm{Im}\left(\tilde{A}_{\sigma_{0}}\tilde{C}_{\sigma_{1}}\right)\cup\mathrm{Im}\left(\tilde{C}_{\sigma_{0}}\right)=\mathbb{R}^n,
\end{array}
\end{align*}
which implies that
$$
\mathrm{rank}\left(\begin{bmatrix}
C_{\sigma_0}\\
C_{\sigma_1}A_{\sigma_0}\\
\vdots\\
C_{\sigma_{T-2}}A_{\sigma_{T-3}}\cdots A_{\sigma_{0}}\\
C_{\sigma_{T-1}}A_{\sigma_{T-2}}\cdots A_{\sigma_{0}}
\end{bmatrix}\right)=n.
$$
This means that with the initial logical state $\alpha$, the system (\ref{eq3.1}) is observable.

Similar to the argument in the proof of Theorem \ref{t4.1}, the initial logical state should also be chosen from $\Delta_N$ arbitrarily. \hfill $\Box$

From the relationship between reachability and controllability of the switched linear systems, we have the following results resorting to the duality principle.
\begin{thm}
\label{t5.2}
The switched linear system (\ref{eq3.1}) is reconstructible under the logically generated switching signal, if and only if there exists a logical input sequence $(\gamma_0,\gamma_1,\cdots,\gamma_{T-1}),~T<\infty$, such that staring from any logical state $\alpha\in[1,N]$, the merged system (\ref{eq5.6}) satisfies
\begin{align}\label{eq5.8}
 {\bf 1}_N^\mathrm{T}\left[\mathrm{Im}(\tilde{G}_{\gamma_{0}}\tilde{G}_{\gamma_{1}}\cdots \tilde{G}_{\gamma_{T-1}})\right]\subset \tilde{\cal R}_T^\alpha(\gamma_{0},\gamma_{1}\cdots,\gamma_{T-1}).
\end{align}
\end{thm}

\begin{rem}
\label{r5.2}
Similar with the argument in Remark \ref{r4.2}, for the merged system (\ref{eq3.5}) whose linear subsystem (\ref{eq3.1}) is reversible, Theorems \ref{t5.1} and \ref{t5.2} are equivalent.
\end{rem}

\section{System Realization Problem}
In the previous section, we investigated four fundamental control properties of discrete-time SLSs where the switching signals are generated by logical control systems. Now it is natural to ponder the following problem: under what condition(s) can the logical system produce a desirable sequence such that the linear system's performance reaches our goal?

A straightforward result is as follows.
\begin{prp}
\label{p6.1}
Consider the logical control network (\ref{eq3.2}). If each state $\theta\in {\cal D}_N$ is one-step controllable from all the states, which in the vector form is
\begin{align}\label{eq6.1}
    \left[L{\bf 1}_M\right]_{\alpha,\beta}>0,\quad \forall\alpha,\beta\in[1,N],
\end{align}
then the four properties of the switched linear system (\ref{eq3.1}) with constrained switching signals are reduced to the unconstrained cases. 
\end{prp}

One should notice that the condition (\ref{eq6.1}) is rather strict for a logical control network. In the following, we will probe some feasible conditions.

We consider two kinds of logical regulating methods:
\begin{itemize}
    \item {\it Case 1:} Generating the switching signal sequences that guarantee the fixed operating times (FOTs) for the subsystems.
    \item {\it Case 2:} Generating the switching signals aligned with a finite reference sequence.
\end{itemize}

\subsection{$\ell$-Step Input-State Set Reachability}
In this subsection, we introduce the $\ell$-step input-state set reachability of logical control networks, which is used hereafter.

Consider the logical control network (\ref{eq3.2.0}) and its ASSR (\ref{eq3.2}). Denote the input set by ${\cal U}:=\{1,2,\cdots,M\}$ and the state set by ${\cal X}:=\{1,2,\cdots,N\}$, then $\Omega\in 2^{{\cal U}\times{\cal X}}\setminus\{\emptyset\}$ is an input-state subset\footnote{${\cal U}\times{\cal X}$ is isomorphic to $\Delta_{MN}$ under logical operations.}. Now we denote $V(\Omega)\in{\cal B}_{MN\times 1}$ the index vector of $\Omega$, which is defined as 
\begin{align*}
    \left[V(\Omega)\right]_i:=\left\{
    \begin{array}{rr}
         1, & i\in \Omega;  \\
         0, & i\notin \Omega.
    \end{array}
    \right.
\end{align*}
One can see that $V(\Omega)=\dsum\limits_{(\gamma,\theta)\in\Omega}\vec{\gamma}\vec{\theta}$.

For a class of initial state subsets $\Omega^0:=\{\Omega_1^0,\Omega_2^0,\cdots \Omega_\alpha^0\}$ and a class of terminal state subsets $\Omega^d:=\{\Omega_1^d,\Omega_2^d,\cdots \Omega_\beta^d\}$ ($\alpha,\beta\in\mathbb{Z}_+$ are the numbers of the subsets), define their index matrices as
\begin{align*}
    P_{\Omega^0}:=\begin{bmatrix}
    V(\Omega_1^0) & V(\Omega_2^0) &\cdots& V(\Omega_\alpha^0)
    \end{bmatrix}\in{\cal B}_{MN\times\alpha},\\
        P_{\Omega^d}:=\begin{bmatrix}
    V(\Omega_1^d) & V(\Omega_2^d) &\cdots& V(\Omega_\beta^d)
    \end{bmatrix}\in{\cal B}_{MN\times\beta}.
\end{align*}

Now we are ready to give the concept of input-state set reachability of logical control networks.
\begin{dfn}\label{d2.2.1}
    Consider the logical control network (\ref{eq3.2}) with a class of initial input-state subsets $\Omega^0=\{\Omega_1^0,\Omega_2^0,\cdots \Omega_\alpha^0\}$ and a class of terminal input-state subsets $\Omega^d=\{\Omega_1^d,\Omega_2^d,\cdots \Omega_\beta^d\}$.
    \begin{enumerate}
        \item The system is $\ell$-step input-state reachable from $(\gamma^0,\theta^0)$ to $(\gamma^d,\theta^d)$ if, there exists at least a logical input sequence $(\gamma(0),\gamma(1),\cdots,\gamma(\ell))$, where $\gamma(0)=\gamma^0$ and $\gamma(\ell)=\gamma^d$, such that $(\gamma^0,\theta^0)$ can be steered to $(\gamma^d,\theta^d)$.
        \item The system is $\ell$-step input-state set reachable from $\Omega_j^0$ to $\Omega_i^d$ if, for some $(\gamma^0,\theta^0)\in\Omega_j^0$ and some $(\gamma^d,\theta^d)\in\Omega_i^d$ there exists at least a logical input sequence $(\gamma(0),\gamma(1),\cdots,\gamma(\ell))$, where $\gamma(0)=\gamma^0$ and $\gamma(\ell)=\gamma^d$, such that the system is reachable from $(\gamma^0,\theta^0)$ to $(\gamma^d,\theta^d)$.
        \item The system is $\ell$-step input-state set reachable at $\Omega_j^0$ if, the system is set reachable from $\Omega_j^0$ to $\forall \Omega_i^d\in\Omega^d$.
        \item $\Omega_i^d$ is global $\ell$-step input-state set reachable if, the system is set reachable from $\forall\Omega_j^0\in\Omega^0$ to $\Omega_i^d$.
        \item The system is $\ell$-step input-state set reachable from $\Omega^0$ to $\Omega^d$ if, for $\forall \Omega_j^0\in\Omega^0$ and $\forall \Omega_i^d\in\Omega^d$, the system is $\ell$-step input-state set reachable from $\Omega_j^0$ to $\Omega_i^d$.
    \end{enumerate}
\end{dfn}

Define the input-state matrix of the logical control network (\ref{eq3.2}) as
\begin{align}
    {\bf L}:={\bf 1}_{M}L={\underbrace{[L^\mathrm{T}~L^\mathrm{T}~\cdots~L^\mathrm{T}]}_{M}}^\mathrm{T}.
\end{align}
Given $\Omega^0$ and $\Omega^d$, we have the $\ell$-step input-state set reachability matrix of the system (\ref{eq3.2}) as 
\begin{align}
    \label{eq2.2.2}
{\cal C}_\ell:=(P_{\Omega}^d)^\mathrm{T}\times_{\cal B}{\bf L}^{(\ell)}\times_{\cal B}P_\Omega^0,
\end{align}
the $\ell$-step input-state set reachability can be verified by the following conditions.
\begin{prp}\label{p2.2.1}
    Consider the logical control network (\ref{eq3.2}) with a group of initial input-state subsets $\{\Omega_1^0,\Omega_2^0,\cdots \Omega_\alpha^0\}$ and terminal input-state subsets $\{\Omega_1^d,\Omega_2^d,\cdots \Omega_\beta^d\}$.
    \begin{enumerate}
        \item The system is $\ell$-step input-state set reachable from $\Omega_j^0$ to $\Omega_i^d$, if and only if
        $[{\cal C}_\ell]_{i,j}=1$.
        \item The system is $\ell$-step input-state set reachable at $\Omega_j^0$, if and only if $\Col_{j}({\cal C}_\ell)={\bf 1}_\beta$.
        \item $\Omega_i^d$ is global $\ell$-step input-state set reachable, if and only if $\Row_{i}({\cal C}_\ell)={\bf 1}_\alpha^\mathrm{T}$.
        \item The system is $\ell$-step input-state set reachable, if and only if ${\cal C}_\ell={\bf 1}_{\beta\times\alpha}$.
    \end{enumerate}
\end{prp}
{\it Proof:}
Given $\gamma(0)=\gamma^0$ and $\theta(0)=\theta^0$, one has
\begin{align*}
\begin{array}{lll}
{\bf L}\times_{\cal B}\vec{\gamma}(0)\vec{\theta}(0)={\bf 1}_{M}\times_{\cal B} L\vec{\gamma}(0)\vec{\theta}(0)\\
=\sideset{}{_{\cal B}}\dsum\limits_{i=1}^{k^m}\delta_{M}^i\vec{\theta}(1):=\vec{\vartheta}(1),
\end{array}
\end{align*}
where $\vec{\vartheta}(t)$ denotes the Boolean sum of all possible input-states $\vec{\gamma}(t)\vec{\theta}(t)$ at time $t$, starting from the initial input-state $\vec{\gamma}(0)\vec{\theta}(0)$. Then one has
\begin{align*}
\begin{array}{rcl}
{\bf L}^{(\ell)}\times_{\cal B}\vec{\gamma}(0)\vec{\theta}(0)&=&{\bf L}^{(\ell-1)}\times_{\cal B}\vec{\vartheta}(1)\\
~&=&{\bf L}^{(\ell-2)}\times_{\cal B}{\bf L}\times_{\cal B}\sideset{}{_{\cal B}}\dsum\limits_{i=1}^{M}\delta_{M}^i\vec{\theta}(1)\\
~&=&{\bf L}^{(\ell-2)}\times_{\cal B}\sideset{}{_{\cal B}}\dsum\limits_{j=1}^{M}\delta_{M}^j\sideset{}{_{\cal B}}\dsum\limits_{i=1}^{M}L\delta_{M}^i\vec{\theta}(1)\\
~&=&{\bf L}^{(\ell-2)}\times_{\cal B}\vec{\vartheta}(2)\\
~&~&\vdots\\
~&=&{\bf L}\times_{\cal B}\vec{\vartheta}(\ell-1)\\
~&=&\vec{\vartheta}(\ell),
\end{array}
\end{align*}
which deduces the following recursion
$$\vec{\vartheta}(t+1)={\bf L}\vec{\vartheta}(t) \quad t\geq 1.$$
Then it can be easily concluded that
\begin{align*}
\left(\vec{\gamma^d}\vec{\theta^d}\right)^\mathrm{T}{\bf L}^{(\ell)}\vec{\gamma}(0)\vec{\theta}(0)
=\left(\vec{\gamma^d}\vec{\theta^d}\right)^\mathrm{T}\vec{\vartheta}(\ell)\\
=\begin{cases}
    1~\Leftrightarrow \mbox{ $(\gamma^d,\theta^d)$ is reachable from $(\gamma^0,\theta^0)$};\\
    0~\Leftrightarrow \mbox{ $(\gamma^d,\theta^d)$ is not reachable from $(\gamma^0,\theta^0)$}.
\end{cases}
\end{align*}
According to the structure of the index matrices, the criteria in Proposition \ref{p2.2.1} are proved.
\hfill $\Box$

\begin{cor}
    \label{c2.2}
    From the proof of Proposition \ref{p2.2.1}, one sees easily that letting the Boolean product be the conventional product and then compute ${\cal C}_\ell$, i.e.,
    $$\tilde{{\cal C}}_\ell:=(P_{\Omega}^d)^\mathrm{T}\times{\bf L}^{\ell}\times P_\Omega^0,$$
    the results in Proposition \ref{p2.2.1} will be quantitative but not qualitative. That is,
    \begin{enumerate}
        \item The number of the $\ell$-length paths that from ${\Omega}_j^0$ to ${\Omega}_i^d$ is $[\tilde{{\cal C}}_\ell]_{i,j}$.
        \item The system is $\ell$-step input-state set reachable at $\Omega_j^0$, if and only if $\Col_j(\tilde{{\cal C}}_\ell)>0$, and each element of $\Col_j(\tilde{{\cal C}}_\ell)>0$ represents the number of the $\ell$-length paths that end in the corresponding input-state sets.
        \item $\Omega_i^d$ is global $\ell$-step input-state set reachable, if and only if $\Row_i(\tilde{{\cal C}}_\ell)>0$, and each element of $\Row_i(\tilde{{\cal C}}_\ell)>0$ represents the number of the $\ell$-length paths that start in the corresponding input-state sets.
        \item The system is $\ell$-step input-state set reachable, if and only if ${\cal C}_\ell>0$.
    \end{enumerate}
\end{cor}

Using the structure matrix $L$, one can draw the input-state dynamic graph of the logical control network (\ref{eq3.2}). An illustrative example is given below to provide an intuitive understanding of input-state set reachability, where the logical control network was originally proposed in \cite{zha10}.

\begin{exa}\label{e2.3}
Consider a logical control network  (\ref{2.1.4}),
where the structure matrix $L=\delta_4[1~1~2~4~4~4~3~3]$. Its input-state dynamic graph is shown in Fig. \ref{fig.1}. 
\begin{figure}[h!]
\centering
\setlength{\unitlength}{0.6 cm}
\begin{picture}(8,6)\thicklines
\put(-0.75,1.5){\oval(3,1)}
\put(3.25,1.5){\oval(3,1)}
\put(7.25,1.5){\oval(3,1)}
\put(-0.75,3.5){\oval(3,1)}
\put(3.25,3.5){\oval(3,1)}
\put(7.25,3.5){\oval(3,1)}
\put(3.25,5.5){\oval(3,1)}
\put(7.25,5.5){\oval(3,1)}
\put(7.25,2){\vector(0,1){1}}
\put(7.25,4){\vector(0,1){1}}
\put(-0.75,3){\vector(0,-1){1}}
\put(3.25,5){\vector(0,-1){1}}
\put(4.75,5.5){\vector(1,0){1}}
\put(4.75,1.5){\vector(1,0){1}}
\put(0.75,1.5){\vector(1,0){1}}
\put(1.75,3.5){\vector(-1,0){1}}
\put(5.75,3.5){\vector(-1,0){1}}
\put(5.75,5.25){\vector(-1,-1){1.25}}
\put(1.75,3.25){\vector(-1,-1){1.25}}
\put(4.75,1.75){\vector(1,1){1.25}}

\put(-2.25,1.5){\line(-1,0){1.25}}
\put(-3.5,1.5){\line(0,1){4}}
\put(-3.5,5.5){\vector(1,0){5.25}}
\put(9.25,5.5){\oval(0.5,0.5)[r]}
\put(8.75,5.75){\line(1,0){0.5}}
\put(9.25,5.25){\vector(-1,0){0.5}}
\put(9.25,1.5){\oval(0.5,0.5)[r]}
\put(8.75,1.75){\line(1,0){0.5}}
\put(9.25,1.25){\vector(-1,0){0.5}}
\put(-2.75,3.5){\oval(0.5,0.5)[l]}
\put(-2.8,3.75){\vector(1,0){0.5}}
\put(-2.75,3.25){\line(1,0){0.5}}

\put(-1.85,1.35){$1\times(2,1)$}
\put(2.15,1.35){$1\times(1,2)$}
\put(6.15,1.35){$1\times(1,1)$}
\put(-1.85,3.35){$2\times(2,1)$}
\put(2.15,3.35){$2\times(2,2)$}
\put(6.15,3.35){$2\times(1,1)$}
\put(2.15,5.35){$2\times(1,2)$}
\put(6.15,5.35){$1\times(2,2)$}
\end{picture}
\caption{Input-state dynamic graph\label{fig.1}}
\end{figure}
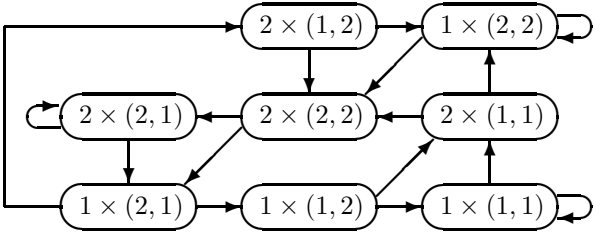
From Fig. \ref{fig.1}, we can observe all the transitions among the input-state variables.

To extract the set reachability, we should first assign the input-state subsets. For instance, let the three layers of nodes be partitioned into three input-state subsets:
\begin{align*}
    \begin{array}{l}
      \Omega_1:=\{2\times(1,2),1\times(2,2)\},\\
      \Omega_2:=\{2\times(2,1),2\times(2,2),2\times(1,1)\},\\
      \Omega_3:=\{1\times(2,1),1\times(1,2),1\times(1,1)\}.
    \end{array}
\end{align*}
{Then the index vectors are}
\begin{align*}
    \begin{array}{l}
      V(\Omega_1)=[0~0~0~1~0~1~0~0]^\mathrm{T},\\
      V(\Omega_2)=[0~0~0~0~1~0~1~1]^\mathrm{T},\\
      V(\Omega_3)=[1~1~1~0~0~0~0~0]^\mathrm{T}.
    \end{array}
\end{align*}
Letting $\Omega^0:=\{\Omega_1\}$ and $\Omega^d:=\{\Omega_2,\Omega_3\}$, we can obtain that
\begin{align*}
    \tilde{{\cal C}}_1=\begin{bmatrix}
        2\\0
    \end{bmatrix}, \quad \tilde{{\cal C}}_2=\begin{bmatrix}
        4\\2
    \end{bmatrix}.
\end{align*}
This result can be observed from Fig. \ref{fig.1} that the nodes on the uppermost layer have two $1$-length paths that can reach the nodes on the middle layer: $(2\times(1,2))\to(2\times(2,2)),~(1\times(2,2))\to(2\times(2,2))$, while the nodes on the middle layer have two $1$-length paths to the nodes on the nethermost layer: $(2\times(2,1))\to(1\times(2,1)),~(2\times(2,2))\to(1\times(2,1))$, but there is no $1$-length path from the uppermost to the nethermost layer.

One can also find the $2$-length paths from Fig. \ref{fig.1} and the number of the paths coincide with the corresponding elements of $\tilde{\cal C}_2$:
\begin{enumerate}
    \item $4$ paths from the uppermost layer to the middle layer: $(2\times(1,2))\to(1\times(2,2))\to(2\times(2,2)),~(2\times(1,2))\to(2\times(2,2))\to(2\times(2,1)),~(1\times(2,2))\to(2\times(2,2))\to(2\times(2,1)),~(1\times(2,2))\to(1\times(2,2))\to(2\times(2,2))$;
    \item $2$ paths from the uppermost layer to the nethermost layer: $(2\times(1,2))\to(2\times(2,2))\to(1\times(2,1)),~(1\times(2,2))\to(2\times(2,2))\to(1\times(2,1))$.
\end{enumerate}

\end{exa}

\subsection{Fixed Operating Time Switching}
In {\it Case 1}, without loss of generality, assume that the FOTs of the corresponding subsystems are positive. Then we have the following conditions.

\begin{thm}
    \label{t6.2}
    Consider system (\ref{eq3.1}) in which the switching signal is generated by the system (\ref{eq3.2}). Denote by $d_1,d_2,\cdots,d_q\in\mathbb{Z}_+\cup\{\infty\}$ the FOTs of the subsystems $\Sigma_1,\Sigma_2,\cdots,\Sigma_q$. The switching signal sequences, which ensure the FOTs, can be generated by the system (\ref{eq3.2}) under some logical inputs, no matter what the initial logical state is, if and only if
\begin{enumerate}
    \item For the subsystem $\Sigma_i$ whose FOT $d_i=1$, there exists at least one logical input $\gamma\in[1,M]$, such that the switching signal $\sigma=i$ can be transferred to other values.
    \item For the subsystem $\Sigma_i$ whose FOT satisfies $1<d_i<\infty$, there exists at least one logical input $\gamma\in[1,M]$, such that the switching signal $\sigma=i$ can be transferred to other values; and exists at least one logical input $\gamma\in[1,M]$, such that the switching signal $\sigma=i$ can keep still.
    \item For the subsystem $\Sigma_i$ whose FOT satisfies $d_i=\infty$, there exists at least one logical input $\gamma\in[1,M]$, such that the switching signal $\sigma=i$ can keep still.
\end{enumerate}
\end{thm}

{\it Proof:} The value of the FOT $d_i$ for the subsystem $\Sigma_i$ can be $d_i=1$, $1<d_i<\infty$, or $d_i=\infty$, and for an initial logical state, the initial switching signal must correspond to the subsystem according with one of the three situations.
\begin{itemize}
    \item For the case of $d_i=1$, once the system is started with $\Sigma_i$ at $t=0$ (or switched to $\Sigma_i$ at time $t=\tau$), the subsystem $\Sigma_i$ should be transferred to the other subsystems immediately at time $t=1$ (or $\tau+1$) using a logical input $\gamma_{1}$ (or $\gamma_{\tau+1}$).\\If there is no such logical input, then it means that once the system is switched to subsystem $\Sigma_i$, the system will always run in this mode, which is forbidden by the FOT.
    \item For the case of $1<d_i<\infty$, once the system is started with $\Sigma_i$ at $t=0$ (or switched to $\Sigma_i$ at time $t=\tau$), the logical system (\ref{eq3.2}) should have the ability to $1)$ let $\sigma=i$ stay if $t<d_i$ (or $t<\tau+d_i$), and $2)$ steer to the other values when $t=d_i$ (or $t=\tau+d_i$). Thus, there should exist at least two logical inputs, such that one can keep $\sigma=i$ still and the other can steer $\sigma$ to other values. Otherwise, the subsystem cannot reach or will exceed the FOT.
    \item For the case of $d_i=\infty$, which means that once the system is started with $\Sigma_i$ at $t=0$ (or switched to subsystem $\Sigma_i$ at time $t=\tau$), it should operate in this mode at $t>0$ (or $t>\tau$). Thus there should exist at least one logical input such that $\sigma=i$ holds. Otherwise, the mode will be changed to other subsystems at $t=1$ (or $t=\tau+1$).
\end{itemize}
The proof is thus completed. \hfill $\Box$

Let ${\cal I}:=\{i\mid d_i=1\}$, ${\cal J}:=\{i\mid 1<d_i<\infty\}$, and ${\cal K}:=\{i\mid d_i=\infty\}$. Define a series of input-state subsets as ${\cal O}_i:=\{\delta_{MN}^\alpha \mid R\delta_{MN}^\alpha=\delta_q^i\}$, and define the singleton version of ${\cal O}_i$ as $\tilde{\cal O}_i:=\{\{\delta_{MN}^\alpha \}\mid R\delta_{MN}^\alpha=\delta_q^i\}$. Now using Proposition \ref{p2.2.1}, we can obtain the following result.

\begin{thm}
\label{t6.3}
The switching signal sequence which guarantees the FOTs of the switched linear system (\ref{eq3.1}) can be generated regardless of the initial logical state of the logical control network (\ref{eq3.2}), if and only if

\begin{enumerate}
    \item $\forall i\in {\cal I}$, the system is $1$-step input-state set reachable from every singleton in $\tilde{\cal O}_i$ to $\Delta_{MN}\setminus{\cal O}_i$, that is,
    \begin{align}\label{eq6.2.1}
       P_{\{\Delta_{MN}\setminus{\cal O}_i\}}^\mathrm{T} \times_{\cal B} {\bf L} \times_{\cal B} P_{\tilde{\cal O}_i}={\bf 1}_{|{\cal O}_i|}^\mathrm{T}. 
    \end{align}
    \item $\forall i\in {\cal J}$, the system is $1$-step input-state set reachable from every singleton in $\tilde{\cal O}_i$ to both $\Delta_{MN}\setminus{\cal O}_i$ and ${\cal O}_i$, that is,
    \begin{align}\label{eq6.2.2}
    \begin{cases}
        P_{\{\Delta_{MN}\setminus{\cal O}_i\}}^\mathrm{T} \times_{\cal B} {\bf L} \times_{\cal B} P_{\tilde{\cal O}_i}={\bf 1}_{|{\cal O}_i|}^\mathrm{T},\\
        P_{\{{\cal O}_i\}}^\mathrm{T} \times_{\cal B} {\bf L} \times_{\cal B} P_{\tilde{\cal O}_i}={\bf 1}_{|{\cal O}_i|}^\mathrm{T}.
    \end{cases}
    \end{align}
    \item $\forall i\in {\cal K}$, the system is $1$-step input-state set reachable from singleton in $\tilde{\cal O}_i$ to ${\cal O}_i$, that is,
    \begin{align}
        \label{eq6.2.3}
        P_{\{{\cal O}_i\}}^\mathrm{T} \times_{\cal B} {\bf L} \times_{\cal B} P_{\tilde{\cal O}_i}={\bf 1}_{|{\cal O}_i|}^\mathrm{T}.
    \end{align}
\end{enumerate}
where ${\bf L}={\bf 1}_M L={\underbrace{[L^\mathrm{T},L^\mathrm{T},\cdots,L^\mathrm{T}]}_M}^\mathrm{T}$ is the input-state matrix of the logical control network (\ref{eq3.2}).
\end{thm}

To avoid frequent switches of the subsystems, the concept of minimum dwell time is usually used when studying switched systems. It can be easily seen that the minimum dwell time is a special case of the FOT. To be specific, we have the following result.
\begin{cor}\label{r6.1}
    If the switching signal sequence is asked to align with a minimum dwell time $\underline{d}_i\in\mathbb{Z}_+$ assigned to each subsystem $\Sigma_i$, then it is obvious that the logical control network (\ref{eq3.2}) has the ability to generate the required switching signal sequences if condition $(2)$ in Theorem \ref{t6.2} (equivalent to condition $(2)$ in Theorem \ref{t6.3}) is satisfied for $\forall i\in[1,q]$.
\end{cor}

\subsection{Finite Reference Signal Switching}
A finite reference signal sequence is given in {\it Case 2}. The key point of this problem is checking whether the required signal sequence can be generated by the logical control network (\ref{eq3.2}). In \cite{zha18}, the authors studied the finite horizon output tracking of Boolean control networks, which provides us with a way to solve the problem. We first generalize the concept of finite horizon output tracking to $\sigma$-tracking of the logical control network (\ref{eq3.2}).

\begin{dfn}\label{d6.1}
Consider the logical control network (\ref{eq3.2}) with an initial state $\theta_0$. The reference signal sequence $(\sigma_0,\sigma_1,\cdots,\sigma_\tau)$ is called trackable if there exists a logical input sequence $\Gamma:=(\gamma_0,\gamma_1,\cdots,\gamma_\tau)$ such that $$\sigma(t,\theta_0,\Gamma)=\sigma_t,\quad t=0,1,\cdots,\tau.$$
\end{dfn}

For a given reference signal sequence $(\sigma_0,\sigma_1,\cdots,\sigma_\tau)$, we define a series of input-state subsets as $${\cal O}_{\sigma_{t}}:=\{\delta_{MN}^{i}\mid R\delta_{MN}^i=\sigma_t\},~t=0,1,\cdots,\tau.$$

Then using the input-state set controllability method and the results in \cite{zha18}, we have the following theorem.
\begin{thm}
    \label{p6.3}
Consider the logical control network (\ref{eq3.2}) with an initial state $\theta_0$. The reference signal sequence $(\sigma_0,\sigma_1,\cdots,\sigma_\tau)$ is trackable, if and only if
\begin{align}
        \label{eq6.3.1}
        P^\mathrm{T}_{\{ {\cal O}_{\sigma_t}\}}{\bf L}\vec{\vartheta}(t-1)>0,\quad t=1,2,\cdots,\tau,
    \end{align}
where $\begin{cases}\begin{array}{l}
        \vec{\vartheta}(0)=\left({\bf 1}_M\vec{\theta}(0) \right)\wedge P_{\{ {\cal O}_{\sigma_0} \}}\\
        \vec{\vartheta}(t)=\left({\bf L}\times_{\cal B}\vec{\vartheta}(t-1)\right)\wedge P_{\{ {\cal O}_{\sigma_{t}}\}}
    \end{array} \end{cases}$.
\end{thm}
{\it Proof:}
It is clear that the tracking problem has a solution if and only if there is an input-state sequence $\{(\gamma_0,\theta_0),(\gamma_1,\theta_1),\cdots,(\gamma_{\tau},\theta_{\tau})\}$, where $\theta_0$ is given, such that $$R\vec{\gamma}_t\vec{\theta}_t=\vec{\sigma}_t,~t=0,1,\cdots,\tau,$$ 
where $\vec{\theta}_{t+1}=L\vec{\gamma}_{t}\vec{\theta}_t,~t=0,1,\cdots,\tau-1$. That is, the input-state sequence satisfies the following condition:
\begin{itemize}
    \item For $t=1,2,\cdots,\tau$, $\vec{\gamma}_{t}\vec{\theta}_{t}$ is $1$-step reachable from $\vec{\gamma}_{t-1}\vec{\theta}_{t-1}$, where $\vec{\gamma}_t\vec{\theta}_t\in {\cal O}_{\sigma_t}$ for $t=0,1,\cdots,\tau$.
\end{itemize}

{\bf Sufficiency:}
Using Proposition \ref{p2.2.1} and its proof, one sees easily that in equation (\ref{eq6.3.1}), $\vec{\vartheta}(t)$ is the sum of all the possible input-states that can generate $\vec{\sigma}_t$ at the $t$-th step. If (\ref{eq6.3.1}) holds, then there is an input-state path such that the above condition is ensured.

{\bf Necessity:} If equation (\ref{eq6.3.1}) does not hold for at least one $t\in [1,\tau]$, then by Proposition \ref{p2.2.1} we know that $\vec{\gamma}_{t}\vec{\theta}_{t}\notin {\cal O}_{\sigma_{t}}$, which contradicts the above condition.
\hfill $\Box$

With the above results, the system realization solvability condition is obtained.

\begin{thm}\label{t6.4}
Consider the switched linear system (\ref{eq3.1}) where the switching signal is generated by the logical control network (\ref{eq3.2}). The signal sequence, which aligns with the reference sequence $(\sigma_1,\sigma_2,\cdots,\sigma_\tau)$, can be generated by the system (\ref{eq3.2}) starting from an initial logical state $\theta_0$, if and only if the reference sequence is $\sigma$-trackable by the system (\ref{eq3.2}).
\end{thm}

\section{Illustrative Example}
\begin{exa}
    \label{e7.1}
    Consider the SLS
\begin{align}\label{eq7.1}
\left\{
\begin{array}{lll}
x(t+1)=A_{\sigma(t)}x(t)+B_{\sigma(t)}u(t),\\
y(t)=C_{\sigma(t)}x(t),
\end{array}\right.
\end{align}
whose switching signals $\sigma(t)=\{1,2\}$ are generated by the following logical control network
\begin{align}\label{eq7.2}
\left\{
\begin{array}{lll}
\vec{\theta}(t+1)=L\ltimes \vec{\gamma}(t)\ltimes \vec{\theta}(t),\\
\vec{\sigma}(t)=R\ltimes \vec{\gamma}(t)\ltimes \vec{\theta}(t),
\end{array}\right.
\end{align}
where
$$A_1=\begin{bmatrix}
    1&2&-1\\0&1&0\\1&-4&3
\end{bmatrix},~B_1=\begin{bmatrix}
    1\\0\\0
\end{bmatrix},~C_1=\begin{bmatrix}
    0&0&1
\end{bmatrix},$$ $$A_2=\begin{bmatrix}
   -2&2&1\\0&-2&0\\1&-4&0
\end{bmatrix},~B_2=\begin{bmatrix}
    0\\1\\0
\end{bmatrix},~C_2=\begin{bmatrix}
    0&1&0
\end{bmatrix},$$
$$L=\delta_{4}[1,1,2,4,4,4,3,3],$$
$$R=\delta_{2}[2,2,1,1,1,2,2,1].$$
With equation (\ref{eq3.4}), one has
\begin{align*}
    {\bf G}=\begin{bmatrix}
G_1,G_2
\end{bmatrix},
\end{align*}
where
\begin{align*}
 G_1=&\left[\begin{smallmatrix}
-2&	2&	1&	-2&	2&	1	&0&	0&	0&	0&	0&	0\\
0&	-2&	0&	0&	-2&	0	&0&	0&	0&	0&	0&	0\\
1&	-4&	0&	1&	-4&	0	&0&	0&	0&	0&	0&	0\\
0&	0&	0&	0&	0	&0&	1	&2&	-1&	0&	0&	0\\
0&	0&	0&	0&	0	&0&	0	&1&	0&	0&	0&	0\\
0&	0&	0&	0&	0	&0&	1 &-4&3& 0&	0&	0\\
0&	0&	0&	0&	0&	0&	0&	0&	0&	0&	0&	0\\
0&	0&	0&	0&	0&	0&	0&	0&	0&	0&	0&	0\\
0&	0&	0&	0&	0&	0&	0&	0&	0&	0&	0&	0\\
0&	0&	0&	0&	0&	0&	0&	0&	0&	1&	2&	-1\\
0&	0&	0&	0&	0&	0&	0&	0&	0&	0&	1&	0\\
0&	0&	0&	0&	0&	0&	0&	0&	0&	1&	-4&	3
\end{smallmatrix}\right]\\
:=&\begin{bmatrix}
    G^1_{1,1} &G^1_{1,2} & {\bf 0} &{\bf 0}\\
    {\bf 0} &{\bf 0} & G^1_{2,3} & {\bf 0}\\
    {\bf 0} &{\bf 0} & {\bf 0} &{\bf 0}\\
    {\bf 0} &{\bf 0} & {\bf 0} &G^1_{4,4}\\
\end{bmatrix},\\
G_2=&\left[\begin{smallmatrix}
0&	0&	0&	0&	0&	0&	0&	0&	0&	0&	0&	0\\
0&	0&	0&	0&	0&	0&	0&	0&	0&	0&	0&	0\\
0&	0&	0&	0&	0&	0&	0&	0&	0&	0&	0&	0\\
0&	0&	0&	0&	0&	0&	0&	0&	0&	0&	0&	0\\
0&	0&	0&	0&	0&	0&	0&	0&	0&	0&	0&	0\\
0&	0&	0&	0&	0&	0&	0&	0&	0&	0&	0&	0\\
0&	0&	0&	0&	0&	0&	-2&	2&	1&	1&	2&	-1\\
0&	0&	0&	0&	0&	0&	0&	-2&	0&	0&	1&	0\\
0&	0&	0&	0&	0&	0&	1&	-4&	0&	1&	-4&	3\\
1&	2&	-1&	-2&	2&	1&	0&	0&	0&	0&	0&	0\\
0&	1&	0&	0&	-2&	0&	0&	0&	0&	0&	0&	0\\
1&	-4&	3&	1&	-4&	0&	0&	0&	0&	0&	0&	0
\end{smallmatrix}\right]\\
:=&\begin{bmatrix}
    {\bf 0} &{\bf 0} & {\bf 0} &{\bf 0}\\
    {\bf 0} &{\bf 0} & {\bf 0} &{\bf 0}\\
    {\bf 0} &{\bf 0} & G^2_{3,3} &G^2_{3,4}\\
    G^2_{4,1} &G^2_{4,2} & {\bf 0} &{\bf 0}\\
\end{bmatrix}.
\end{align*}

We first check the control attractors, as well as their attractor basins, of the logical network (\ref{eq7.2}). Using Proposition 7 in \cite{zha19}, we know that all the logical states are control attractors and the attract basin of $\delta_4^4$ is the whole state space $\Delta_4$. Hence it is only needed to check if equations (\ref{eq4.1}), (\ref{eq4.2}), (\ref{eq5.7}), and (\ref{eq5.8}) hold for $\alpha=4$. 

Using Algorithm \ref{a4.3}, we have
\begin{align*}
    {\cal G}_3=\begin{bmatrix}
   *  &G^1_{1,2}G^1_{2,3}G^2_{3,4}\\
   *  &G^1_{2,3}G^2_{3,3}G^2_{3,4}+G^1_{2,3}G^2_{3,4}G^1_{4,4}\\
   *  &G^2_{3,3}G^2_{3,3}G^2_{3,4}+G^2_{3,3}G^2_{3,4}G^1_{4,4}+G^2_{3,4}G^1_{4,4}G^1_{4,4}\\
   *  &G^2_{4,2}G^1_{2,3}G^2_{3,4}+G^1_{4,4}G^1_{4,4}G^1_{4,4}\\
\end{bmatrix},
\end{align*}
where the first three columns of blocks are omitted because only the $4$-th column of the block matters.

It can be found that the equations (\ref{eq4.1}) and (\ref{eq4.2}) hold for logical input sequences 
$$(2,2,2),~(2,2,1),~(2,1,2),~(2,1,1),~(1,2,2).$$
Thus, the SLS is reachable and controllable.

Using a similar computation process, one can also conclude that the system is observable and reconstructible: with $\alpha=4$, equations (\ref{eq5.7}), and (\ref{eq5.8}) hold for all the $3$-length logical input sequences except $(2,2,2)$.
\end{exa}

\section{Conclusion}

This paper presents a novel approach to studying the control properties of SLSs with logic dynamic switching. The proposed approach combines the ASSR method and merged hybrid systems to analyze the reachability, controllability, observability, and reconstructibility of such systems. Additionally, two types of system realization problems are investigated by introducing the concept of input-state set reachability. The proposed methods can be extended to continuous-time linear systems.

\appendix
\section*{Appendix}
\section{Semi-Tensor Product of Matrices}
We briefly introduce the mathematical tool used in this paper, that is, the semi-tensor product (STP) of matrices and the algebraic state-space representation (ASSR) method of logical control networks.
\begin{dfn}(\cite{che11})\label{d2.1.1} Let ~$A\in {\cal M}_{m\times n}$,  $B\in {\cal M}_{p\times q}$, and the least common multiple of $n$ and $p$ be $t=\lcm(n,p)$ (the least common multiple of $n$ and $p$).
The STP of $A$ and $B$, denoted by $A\ltimes B$, is defined as
\begin{equation} \left(A\otimes I_{t/n}\right)\left(B\otimes I_{t/p}\right).
\end{equation}
\end{dfn}

From Definition \ref{d2.1.1}, one sees that the STP is a generalization of the conventional matrix product because they are equivalent when $n=p$. Hence, we do not need to deliberately distinct them, and in this paper, if it will not cause misunderstanding, the symbol ``$\ltimes$" is omitted.

Consider an $n$-node $\kappa$-value logical control system whose logical evolutionary dynamics is
\begin{align}\label{2.1.1}
\begin{cases}
\theta_1(t+1)=f_1(\theta_1(t),\cdots,\theta_n(t);\gamma_{_1}(t),\cdots,\gamma_{m}(t)),\\
\theta_2(t+1)=f_2(\theta_1(t),\cdots,\theta_n(t);\gamma_{_1}(t),\cdots,\gamma_{m}(t)),\\
~\quad\quad\quad\quad\vdots\\
\theta_n(t+1)=f_n(\theta_1(t),\cdots,\theta_n(t);\gamma_{_1}(t),\cdots,\gamma_{m}(t)),
\end{cases}
\end{align}
where $\theta_i(t)\in {\cal D}_\kappa,~i\in[1,n]$ are the state nodes, $\gamma_j(t)\in {\cal D}_\kappa,~j\in[1,m]$ are the input nodes, $f_i:{\cal D}_{\kappa}^{m+n}\rightarrow {\cal D}_{\kappa}$ are logical functions.

Identifying $\alpha\sim \d_\kappa^{\alpha}$ and $\beta\sim\delta_{\kappa}^\beta$ for $\alpha,\beta\in[1,\kappa]$\footnote{For $2$-value logical networks (i.e. Boolean networks), the domain is usually defined as $\{1,0\}$, which is equivalently defined as $\{1,2\}$ in this paper, just for simplicity of expression.}, $\theta_i$ and $\gamma_j$ can be expressed into their vector forms as
\begin{align*}
    \vec{\theta}_i:=\delta_{\kappa}^\alpha,\quad i\in[1,n];\\
    \vec{\theta}_j:=\delta_{\kappa}^\beta,\quad j\in[1,m].
\end{align*}

Denote by $\theta=(\theta_1,\theta_2,\cdots,\theta_n)\in {\cal D}_\kappa^n$ and $\gamma=(\gamma_1,\gamma_2,\cdots,\gamma_m)\in {\cal D}_\kappa^m$ the overall state variable and overall input variable. Their vector form expressions are
\begin{align*}
\vec{\theta}:=\ltimes_{i=1}^n\vec{\theta}_i\in\Delta_{\kappa^n},\\
\vec{\gamma}:=\ltimes_{j=1}^m\vec{\gamma}_j\in\Delta_{\kappa^m}.
\end{align*}

The following proposition is borrowed from \cite{che11} and \cite{che12}.
\begin{prp}\label{p2.1.1}
\begin{itemize}
\item[(i)] Let $f:{\cal D}_\kappa^{m+n}\ra {\cal D}_\kappa$, expressed by $\zeta=f(\theta_1,\theta_2,\cdots,\theta_n;\gamma_1,\gamma_2,\cdots,\gamma_m)$, be a logical function. Then there exists a unique logical matrix $M_f\in {\cal L}_{\kappa\times \kappa^{m+n}}$, called the structure matrix of $f$, such that in vector form the logical function can be expressed by
\begin{align}\label{2.1.3}
\vec{\zeta}=M_f\vec{\gamma}\vec{\theta}.
\end{align}
\item[(ii)] Let $M_i$ be the structure matrix of the logical function $f_i$, $i=1,2,\cdots,n$. Then there exists a unique logical matrix $M\in {\cal L}_{\kappa^n\times \kappa^{m+n}}$ such that in vector form, logical control network (\ref{2.1.1}) can be expressed by
\begin{align}\label{2.1.4}
\vec{\theta}(t+1)=L\vec{\gamma}(t)\vec{\theta}(t),
\end{align}
where $L=M_1*M_2*\cdots*M_n$ is called the structure matrix of the logical control network (\ref{2.1.1}).
\end{itemize}
Equation (\ref{2.1.4}) is called the ASSR of the logical control network (\ref{2.1.1}).
\end{prp}

\begin{dfn}\cite{zha19}\label{d3.1}
Consider the logical system (\ref{eq3.2}). Given a state subset $V\subset \Delta_N$:
\begin{enumerate}
\item A state $\vec{\theta}^*\in V$ is a control fixed point in $V$, if there exists an input $\vec{\gamma}\in\Delta_M$ such that $L\vec{\gamma}\vec{\theta}^*=\vec{\theta}^*$.
\item A state subset $\{{\theta}(0),{\theta}(1),\cdots,{\theta}(\ell-1)\}\subset{{\cal D}_N}$ is called a control cycle in $V$, if there exists an input sequence $(\gamma(0),\gamma(1),\cdots,\gamma(\ell-1))$ such that
\begin{itemize}
    \item $\vec{\theta}(t+1)=L\vec{\gamma}(t)\vec{\theta}(t),\quad t\in[0,\ell-2]$;
    \item $\vec{\theta}(0)=L\vec{\gamma}({\ell-1})\vec{\theta}({\ell-1})$.
\end{itemize}
\item The control fixed points and control cycles are collectively referred to as the control attractors; the union of all the control fixed points and control cycles in $V$ is called the control attractor set of $V$.
\item For a control attractor, its attract basin is the set of all the states that can be steered to the control attractor.
\end{enumerate}
\end{dfn}

\end{document}